\documentclass[%
 aip,
 amsmath,amssymb,
 reprint,%
cha]{revtex4-1}

\usepackage{graphicx}% Include figure files
\usepackage{dcolumn}% Align table columns on decimal point
\usepackage{bm}% bold math
\usepackage[utf8]{inputenc}
\usepackage[T1]{fontenc}
\usepackage{mathptmx}

 \usepackage{hyperref}
 \hypersetup{colorlinks=true,urlcolor=blue}

\begin{document}

% \preprint{AIP/123-QED}

\title{Comment on ``The Winfree model with non-infinitesimal phase-response curve: Ott-Antonsen theory'' [Chaos {\bf 30}, 073139 (2020)]}
% Force line breaks with \\

\author{Diego Paz\'o}
 \affiliation{Instituto de F\'isica de Cantabria (IFCA), CSIC-Universidad de
 Cantabria, 39005 Santander, Spain}%Lines break automatically or can be forced with \\
\author{Rafael Gallego}%
% \email{rgallego@uniovi.es}
\affiliation{Departamento de Matem\'aticas, Universidad de Oviedo, Campus de
Viesques, 33203 Gij\'on, Spain}%

             %  but any date may be explicitly specified

\begin{abstract}
{\em This article may be downloaded for personal use only. Any other use requires prior permission of the author and AIP Publishing. 
This article appeared in Chaos {\bf 31}, 018101 (2021) and may be found 
\href{https://aip.scitation.org/doi/full/10.1063/5.0036357}{here}}\\

% It is always \today, today,
In a recent paper [Chaos {\bf 30}, 073139 (2020)] we analyzed an extension of the Winfree model
with nonlinear interactions.
The nonlinear coupling function $Q$ was mistakenly identified with the non-infinitesimal
phase-response curve (PRC). Here, we asses to what extent $Q$ and  the actual PRC differ in practice.
By means of numerical simulations, we compute the PRCs corresponding to the $Q$ functions previously considered. 
The results confirm a qualitative similarity between the PRC and the coupling function $Q$ in all cases. 
% \href{http://dx.doi.org/10.1063/5.0015131}{doi: 10.1063/5.0015131}
\end{abstract}

\maketitle

In Ref.~\cite{PG20} we studied this generalization of the Winfree model of 
globally coupled phase oscillators:
\begin{subequations} \label{winfree} 
\begin{eqnarray}
\dot\theta_i&=&\omega_i+ Q(\theta_i,A), \label{wina}\\
A&=& \frac{\epsilon}{N} \sum_{j=1}^N P(\theta_j). \label{winb}
\end{eqnarray}
\end{subequations}
 Here, $A$ is proportional to the sum over the pulses emitted by the $N$ oscillators of the population.
In contrast to the original model \cite{Win67,Win80}, function $Q$ in Eq.~\eqref{wina} has a nonlinear dependence on 
the mean field $A$. The motivation for this is the fact that nonlinearity is an unavoidable consequence 
of applying phase reduction beyond the first order to oscillator ensembles
\cite{RP19}. Note that a Taylor expansion of $Q$ to $n$th order in $A$ yields up to $(n+1)$-body  
phase interactions, similarly to Ref.~\cite{leon19}.

We mistakenly called $Q$ `non-infinitesimal phase-response 
curve' in Ref.~\cite{PG20}. Properly speaking, function $Q$ is 
a non-linear `coupling function' \cite{RP19}.
The aim of this comment is to clarify to what extent the coupling function $Q$
determines the actual phase-response curve (PRC). The PRC quantifies the phase shift
gained by an oscillator in response to an external stimulus \cite{Izh07}.
There is no analytic relation between $Q$ and the PRC
beyond the small $\epsilon$ limit; in that case $Q(\theta,A)\simeq\tilde Q(\theta) A$, 
where $\tilde Q$ turns out to be so-called infinitesimal PRC (iPRC). 
In consequence, we rely here on numerical simulations to compute the PRC empirically. 

The family of functions $Q(\theta,A)$ 
considered in \cite{PG20} was:
\begin{equation}
 Q(\theta,A)=f_1(A) (1-\cos\theta) - f_2(A) \sin\theta.
\label{Q}
 \end{equation}
Four representative pairs of functions  $f_{1,2}(A)$ were studied in detail in \cite{PG20}
and the corresponding coupling functions $Q(\theta,A)$
were depicted in 
Fig.~2 of Ref.~\cite{PG20}. With the aim of comparing them, we obtain the PRC for each of
the four coupling functions $Q$ considered in \cite{PG20}.

The PRC value depends on the timing as well as on the specific shape of the stimulus, 
which is not necessarily weak or brief ~\cite{Izh07}. 
Numerically, we obtain the PRC
measuring the effect on one oscillator's phase of a pulse generated by another oscillator.
This means that the two oscillators are unidirectionally coupled (i.e., a master-slave configuration).
We adopt $\omega=1$ as the natural frequency for both, perturbed and perturbing oscillators,
which is the obvious choice as it is the central frequency 
of the distribution in \cite{PG20}.
Moreover, we follow \cite{PG20} and use
the same $2\pi$-periodic symmetric unimodal pulse function $P(\theta)$.
It vanishes at $\theta=\pm\pi$, and
a free parameter $r<1$ controls the 
narrowness of $P$: The height of the pulse is $P(0)=2/(1-r)$, and $\lim_{r\to1}P(\theta)=2\pi\delta(\theta)$.
In this comment we consider two different pulse widths: $r=0.9$ 
(the value selected in \cite{PG20}), and $r=0.99$ 
corresponding to an extremely narrow pulse. 

\begin{figure*}
  \begin{minipage}[c]{0.65\textwidth}
    \includegraphics[width=\textwidth]{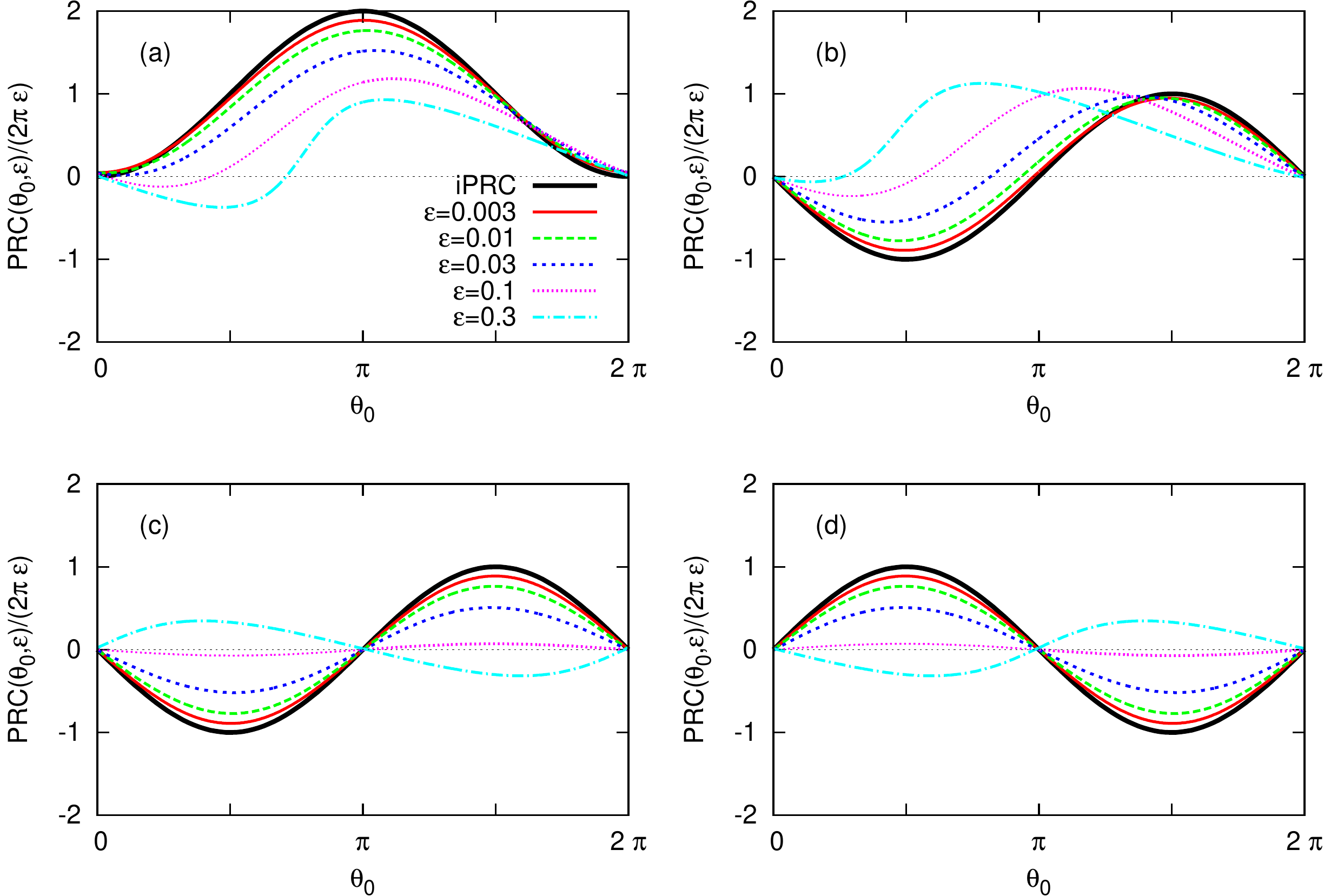}
  \end{minipage}\hfill
  \begin{minipage}[c]{0.3\textwidth}
    \caption{PRCs for cases (a-d) in Ref.~\cite{PG20}. The set of five $\epsilon$ values used in indicated in panel (a). 
    The coupling function \eqref{Q}
in each panel is:    
(a) $f_1=A/(1+A)=f_2/A$; (b) $f_1=A^2/(1+A)=A f_2$;
(c) $f_1=0$, $f_2=A(1-A)/(1+A)$; (d) $f_1=0$, $f_2=A(A-1)/(1+A)$.  
    The pulse acting on the oscillator is
$P(t-\pi)=\frac{(1-r)[1+\cos(t-\pi)]}{1-2r\cos(t-\pi) +r^2}$, with $r=0.9$.} 
% \label{}
  \end{minipage}
\end{figure*}

The simulation starts at time $t=0$ with
the (slave) oscillator at an initial phase $\theta_{\rm in}$. Then,
we let it to evolve under the influence of
the forcing oscillator. The phase of this one grows linearly, such that the input felt by the first oscillator is $A(t)=\epsilon P(t-\pi)$. 
Parameter $\epsilon$ determines the strength of the stimulus.
The simulation runs from $t=0$ to $t=2\pi$, since $A$ exactly vanishes at these times. 
Note that
we do not need to run the simulation further since phase oscillators are governed by first-order differential equations.
For a given $\epsilon$ value, we measure the phase shift at $t=2\pi$
such that $\mathrm{PRC}(\theta_0,\epsilon)=\theta(t=2\pi)-(\theta_{\rm in}+2\pi)$.
The phase $\theta_0$ in the argument of the PRC is the phase value when $A$
attains its maximum, assuming no input exists: $\theta_0=\theta_{\rm in}+\pi$.
The results are shown in Figs.~1 and 2 for a set of $\epsilon$ values; in each panel for one 
particular coupling function $Q(\theta,A)$ already adopted in \cite{PG20}.
In all panels, the corresponding iPRC is shown as a reference.
Note that the normalization of the $y$-axis in Figs.~1 and 2 includes a $2\pi$ factor 
---in addition to $\epsilon$--- because
this is the integral of the pulse over an interval of length $2\pi$. 
Figures 1 and 2 are quite similar, though  for $r=0.9$ (Fig.~1) the PRCs 
remain closer to their 
iPRCs up to a larger $\epsilon$ value.

 \begin{figure*}
  \begin{minipage}[c]{0.65\textwidth}
    \includegraphics[width=\textwidth]{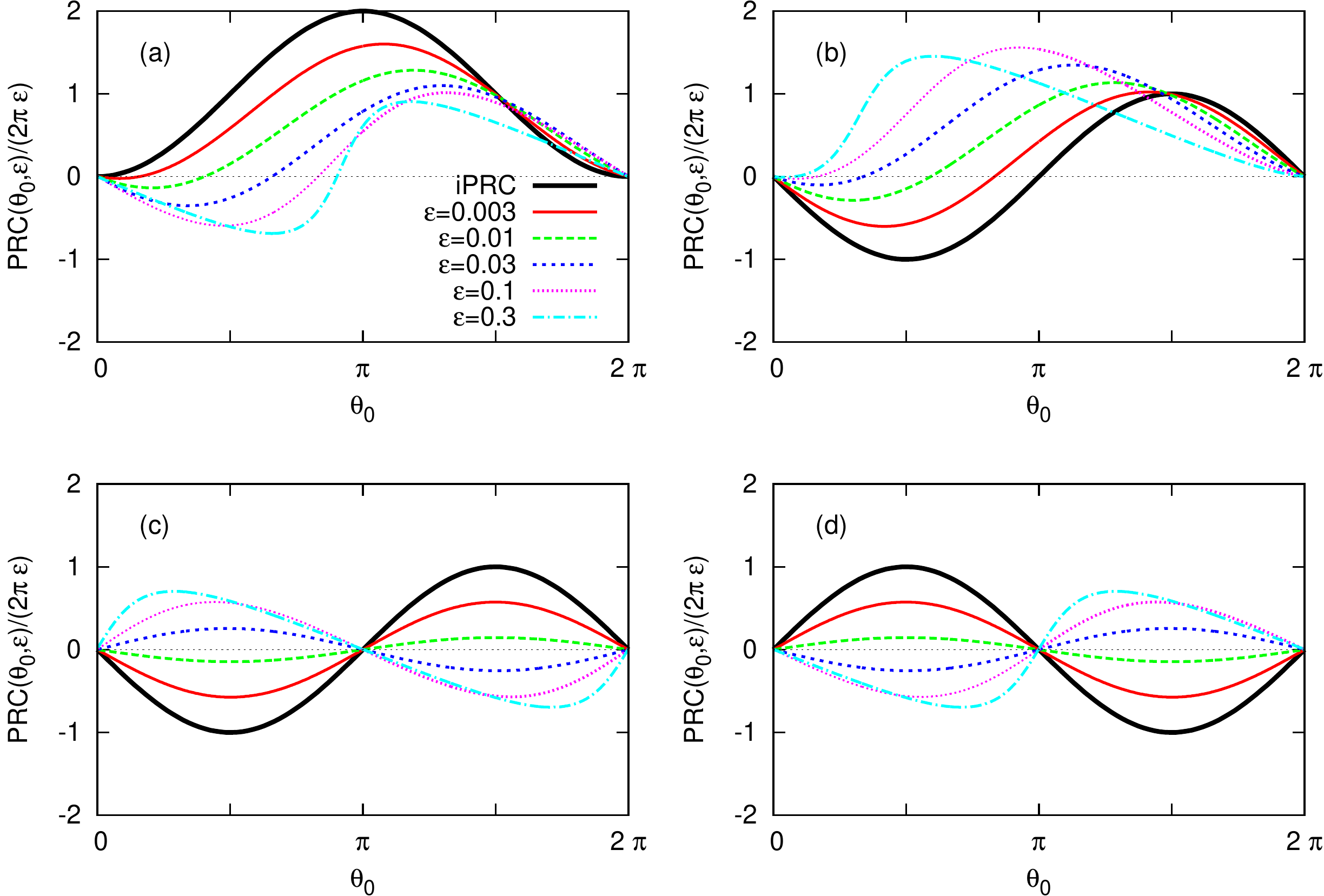}
  \end{minipage}\hfill
  \begin{minipage}[c]{0.3\textwidth}
    \caption{The same as Fig.~1 with $r=0.99$.} 
% \label{}
  \end{minipage}
\end{figure*}

The comparison of the PRCs in this comment with the corresponding $Q$ functions in Fig.~2 of Ref.~\cite{PG20} 
evidences that $Q(\theta,A)$ is not simply the PRC.
Indeed, $Q$ in \eqref{Q} has only the first harmonic in $\theta$, whereas the non-infinitesimal PRCs
in the figures display additional Fourier components.
In spite of these dissimilarities, simple visual inspection indicates that 
the PRC strongly resembles the coupling function $Q$ in all four cases. For example,
we observe the same loss of non-negativeness of the (type-I) iPRC as $A$ increases
in panel (a), or the
transition from a synchronizing iPRC to a desynchronizing PRC for large enough $A$ in panel (c).
Summarizing, our simulations confirm that  the main attributes of the coupling function $Q$ 
are shared by the non-infinitesimal PRC.

 \begin{acknowledgments}
We acknowledge support by the Agencia Estatal de Investigaci\'on and 
Fondo Europeo de Desarrollo Regional under Project No.~FIS2016-74957-P (AEI/FEDER, EU).
 \end{acknowledgments}

% \section*{DATA AVAILABILITY}
%  
%  The data that support the findings of this study are available from the corresponding author 
%  upon reasonable request.

\section*{REFERENCES}
%  \bibliography{bibliografia}% Produces the bibliography via BibTeX.

%merlin.mbs aipnum4-1.bst 2010-07-25 4.21a (PWD, AO, DPC) hacked
%Control: key (0)
%Control: author (8) initials jnrlst
%Control: editor formatted (1) identically to author
%Control: production of article title (0) allowed
%Control: page (1) range
%Control: year (1) truncated
%Control: production of eprint (0) enabled
%

\end{document}